\documentclass[3p,times,procedia]{elsarticle}
\usepackage{nupha_ecrc}

\usepackage{xspace}

\volume{00}

\firstpage{1}

\journalname{Nuclear Physics A}

\runauth{Jochen Th\"{a}der}

\jid{nupha}

\jnltitlelogo{Nuclear Physics A}

\usepackage{amssymb}

\usepackage{lineno}

\usepackage[figuresright]{rotating}

\newcommand{\KV}{\ensuremath{\kappa\sigma^2}\xspace}
\newcommand{\SD}{\ensuremath{S\sigma}\xspace}
\newcommand{\SDK}{\ensuremath{S\sigma/{\rm Skellam}}\xspace}
\newcommand{\VM}{\ensuremath{\sigma^2/M}\xspace}

\newcommand{\GeV}{GeV\xspace}
\newcommand{\MeVc}{\ensuremath{\mathrm{MeV}\kern-0.05em/\kern-0.02em c}\xspace}
\newcommand{\GeVc}{\ensuremath{\mathrm{GeV}\kern-0.05em/\kern-0.02em c}\xspace}
\newcommand{\GeVcSq}{\ensuremath{\mathrm{GeV}\kern-0.05em/\kern-0.02em c^2}\xspace}

\newcommand{\sqrtSnn}{\ensuremath{\sqrt{s_{\mathrm{NN}}}}\xspace}

\newcommand{\sqrtSnnE}[2][GeV]{$\sqrtSnn = #2\,\mathrm{#1}$\xspace}
\newcommand{\sqrtSnnBES}[3][GeV]{$\sqrtSnn = #2  \;\mathrm{to}\;  #3\,\mathrm{#1}$\xspace}

\newcommand{\snn}{\sqrtSnn}

\newcommand{\mub}{\ensuremath{\mu_{\mathrm{B}}}\xspace}

\newcommand{\pt}{\ensuremath{p_{\mathrm{T}}}\xspace}
\newcommand{\ptRange}[2]{\ensuremath{#1\!<\!\unit[\pt]{(\GeVc)}\!<\!#2}\xspace}

\newcommand{\deta}{\ensuremath{\Delta\eta}\xspace}

\newcommand{\dEdx}{\ensuremath{\mathrm{d}E/\mathrm{d}x}\xspace}

\newcommand{\meanNpart}{\ensuremath{\langle N_{\mathrm{part}} \rangle}\xspace}

\newcommand{\netQ}{\ensuremath{\Delta N_{\rm Ch}}\xspace}
\newcommand{\netK}{\ensuremath{\Delta N_{\rm K}}\xspace}
\newcommand{\netP}{\ensuremath{\Delta N_{\rm P}}\xspace}

\DeclareRobustCommand{\unit}[2][]{%
        \begingroup%
                \def\0{#1}%
                \expandafter%
        \endgroup%
        \ifx\0\@empty%
                \ensuremath{\mathrm{#2}}%
        \else%
                \ensuremath{#1\,\mathrm{#2}}%
        \fi%
        }
        
\DeclareRobustCommand{\unitfrac}[3][]{%
        \begingroup%
                \def\0{#1}%
                \expandafter%
        \endgroup%
        \ifx\0\@empty%
                \raisebox{0.98ex}{\ensuremath{\mathrm{\scriptstyle#2}}}%
                \nobreak\hspace{-0.15em}\ensuremath{/}\nobreak\hspace{-0.12em}%
                \raisebox{-0.58ex}{\ensuremath{\mathrm{\scriptstyle#3}}}%
        \else
                \ensuremath{#1}\,%
                \raisebox{0.98ex}{\ensuremath{\mathrm{\scriptstyle#2}}}%
                \nobreak\hspace{-0.15em}\ensuremath{/}\nobreak\hspace{-0.12em}%
                \raisebox{-0.58ex}{\ensuremath{\mathrm{\scriptstyle#3}}}%
        \fi%
}

\begin{document}

\begin{frontmatter}





\dochead{}

\title{Higher Moments of Net-Particle Multiplicity Distributions}
\author{Jochen Th\"{a}der  (for the STAR Collaboration\footnote{A list of members of the STAR Collaboration and acknowledgements can be found at the end of this issue.})}
\address{Nuclear Science Division, Lawrence Berkeley National Laboratory, Berkeley, CA 94720, USA}
\ead{jmthader@lbl.gov}

\begin{abstract}
Studying fluctuations of conserved quantities, such as baryon number, strangeness, and charge, provides insights into the properties of matter created in high-energy nuclear collisions. Lattice QCD calculations suggest that higher moments of these quantities are sensitive to the phase structure of the hot and dense nuclear matter created in such collisions. 
In this paper, we present first experimental results of volume and temperature independent cumulant ratios of net-charge and net-proton distributions in Au+Au collisions at \sqrtSnnE{14.5} completing the first RHIC Beam Energy Scan (BES-I) program for \sqrtSnnBES{7.7}{200}, together with the first measurement of fully corrected net-kaon results, measured with the STAR detector at RHIC at mid-rapidity and a transverse momentum up to \pt = \unit[2]{\GeVc}. The pseudo-rapidity dependence of the \sqrtSnnE{14.5} net-charge cumulant ratios is discussed. The estimated uncertainties on the ratio $c_4/c_2$, the most statistics-hungry of the present observables,  at \sqrtSnnE{7.7}  in the upcoming RHIC BES-II program will also be presented.

\end{abstract}


\end{frontmatter}



\section{Introduction}
The first RHIC Beam Energy Scan (BES-I) program allows one to map the QCD phase diagram, varying collision energy \sqrtSnnBES{7.7}{200} of Au+Au collisions and, thereby, the baryon chemical potential \mub and temperature $T$. There is a smooth cross over at a vanishing \mub, while at higher \mub, model calculations suggest the existence of a first-order phase transition. Therefore, thermodynamic principles suggest that there should be a critical point in the QCD phase diagram, where the first-order phase transition ends and the transition becomes a cross-over \cite{Aoki2006}.

Fluctuations in event-by-event multiplicity distributions of conserved quantities such as net-charge \netQ, net-strangeness (proxy: net-kaon \netK), and net-baryon number (proxy: net-proton \netP)	would indicate a critical behavior. The moments of these distributions are proportional to powers of the correlation length $\xi$, with increasing sensitivity for higher order moments ($\langle (\Delta N)^3 \rangle \propto \xi^{4.5}$, $\langle (\Delta N)^4 \rangle \propto \xi^7$) \cite{PhysRevLett.102.032301, PhysRevLett.107.052301, PhysRevLett.103.262301}. Furthermore, the cumulants $c_{\rm i}$ of these distributions are expected to  be linked to susceptibilities $\chi_{\rm i}$ \cite{PhysRevD.79.074505}, which can be obtained from QCD model calculations \cite{Gavai:2011aa, Gupta24062011}. This allows for a direct comparison of experimentally measured volume independent cumulant ratios ($c_2/c_1 = \VM$, $c_3/c_2 = \SD$, and $c_4/c_2 = \KV$) and theoretically obtained susceptibility ratios, where the volume and temperature dependent terms cancel.
In the absence of a critical point, the hadron resonance gas model \cite{Garg:2013aa} suggests that the \KV values will be close to unity and have a monotonic dependence on \snn \cite{Luo_VolumeFluct} following the Poisson expectation.


\section{Analysis Details}
The STAR (Solenoidal Tracker At RHIC) detector at Brookhaven National Laboratory has a large uniform acceptance at mid-rapidity and excellent particle identification capabilities. The main detectors used in these analyses are the Time Projection Chamber (TPC) \cite{Anderson:2003aa} and the Time-Of-Flight detector (TOF) \cite{Llope:2012aa}. As the main tracking device, the TPC provides full azimuthal acceptance for tracks in the pseudo-rapidity region $|\eta|\!<\!1$. In addition, it provides charged particle identification via the measurement of the specific energy loss \dEdx. The TOF detector provides a similar acceptance as the TPC and its velocity information is used for particle identification via the mass-squared, $m^2$.
The analyses have been carried out event-by-event using minimum-bias events, rejecting piled-up and other background events such as beam-pipe interactions using the TOF information and other global observables. Only events with a reconstructed primary vertex position in the fiducial region $|v_z|\!<\!\unit[30]{cm}$ ($<\!\unit[50]{cm}$ for \unit[7.7]{\GeV}) and $|v_r|\!<\!\unit[1]{cm}$ were considered. All tracks are required to have a minimum length of 20 hits in the TPC to allow for a good two-track separation. In order to reduce the contamination from secondary charged particles, only primary particles have been selected, requiring a distance of closest approach (DCA) to the primary vertex of less than \unit[1]{cm}.

\begin{figure}[htbp]
\begin{center}
\vspace*{-3mm}
\includegraphics[scale=0.75]{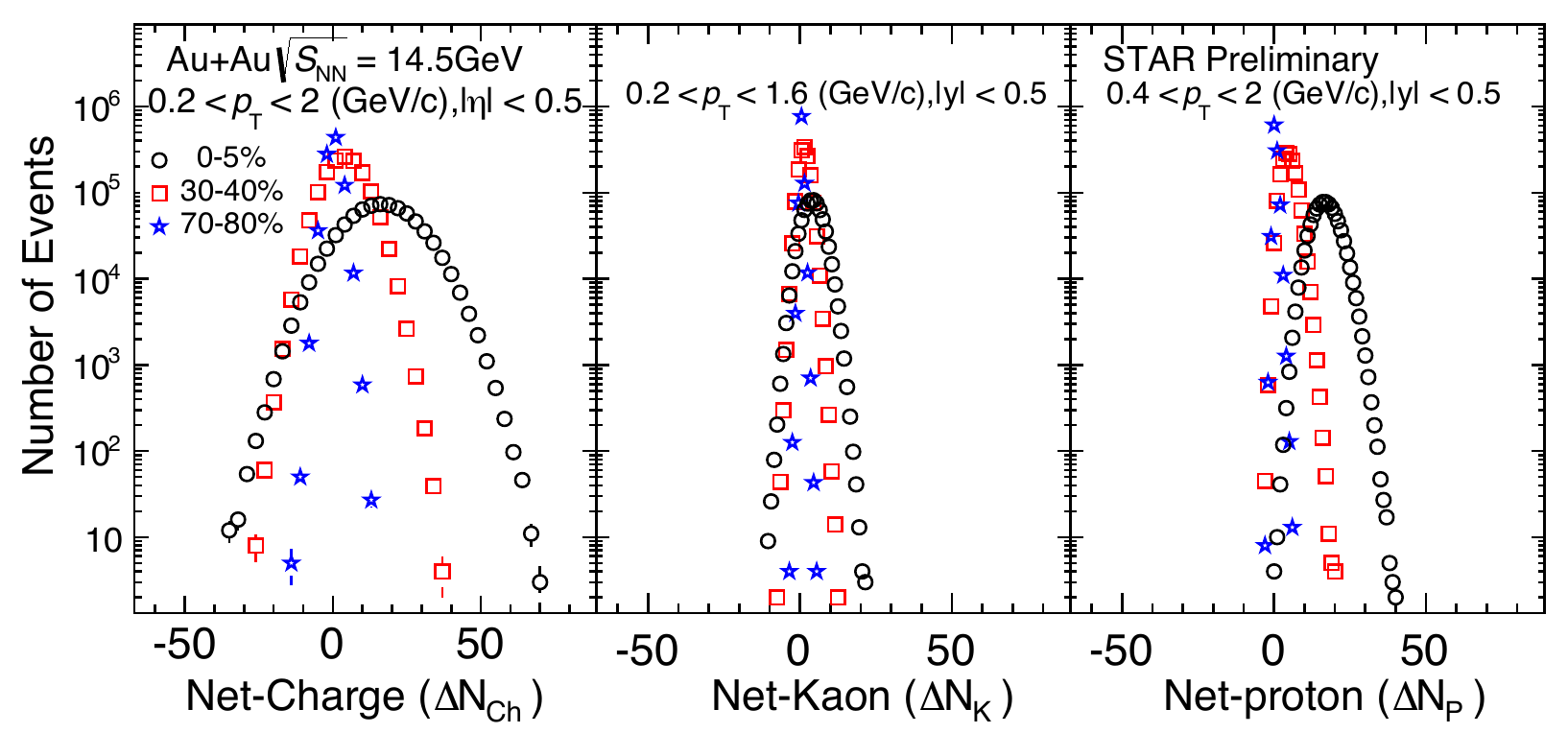}
\vspace*{-4mm}
\caption{Uncorrected raw event-by-event net-particle multiplicity distributions for Au+Au collisions at \sqrtSnnE{14.5} for \netQ (left panel), \netK (middle panel) and \netP (right panel) for 0-5\% top central (black circles),  30-40\% central (red squares), and 70-80\% peripheral collisions (blue stars).} 
\label{unCorrDist}
\vspace*{-5mm}
\end{center}
\end{figure}

The net-particle quantities are formed event-by-event as, $\netQ\!=\!N_{\rm pos}\!-\!N_{\rm neg}$, $\netK\!=\!N_{K^{+}}\!-\!N_{K^{-}}$, and $\netP\!=\!N_{p}\!-\!N_{\overline{p}}$. The measurements of the identified particles have 	been carried out within the rapidity range of $|y|\!<\!0.5$ and in the transverse momentum range of \ptRange{0.2}{1.6} for kaons and \ptRange{0.4}{2.0} for protons. The kaons (protons) have been identified using only the TPC \dEdx information below $\pt\!<\!\unit[0.4]{\GeVc}$ ($\pt\!<\!\unit[0.8]{\GeVc}$) and a combination of TPC and TOF information above. Charged particles have been measured within the pseudo-rapidity range of $|\eta|\!<\!0.5$ and \ptRange{0.2}{2.0}, while the protons below $\pt\!<\!\unit[0.4]{\GeVc}$ have been rejected to reduce the influence of spallation protons.
The centrality classes are bin-width corrected values \cite{Luo_cbwc} from Glauber model fits to the the total charged particle multiplicity distribution ($0.5\!<\!|\eta|\!<\!1.0$), except for net-kaons (net-protons) for which the total multiplicity of pions and protons (pions and kaons) within $|\eta|\!<\!1.0$ was used.
%
%
For illustration purposes only, Fig.~\ref{unCorrDist} shows the uncorrected event-by-event net-particle multiplicity distributions for Au+Au collisions at \sqrtSnnE{14.5} for \netQ, \netK, and \netP in three centrality intervals. The widest distribution is observed for the \netQ and the narrowest for \netK. 

\section{Results and Discussion}

\begin{figure}[tbp]
\begin{center}
\vspace*{-3mm}
\includegraphics[scale=0.30]{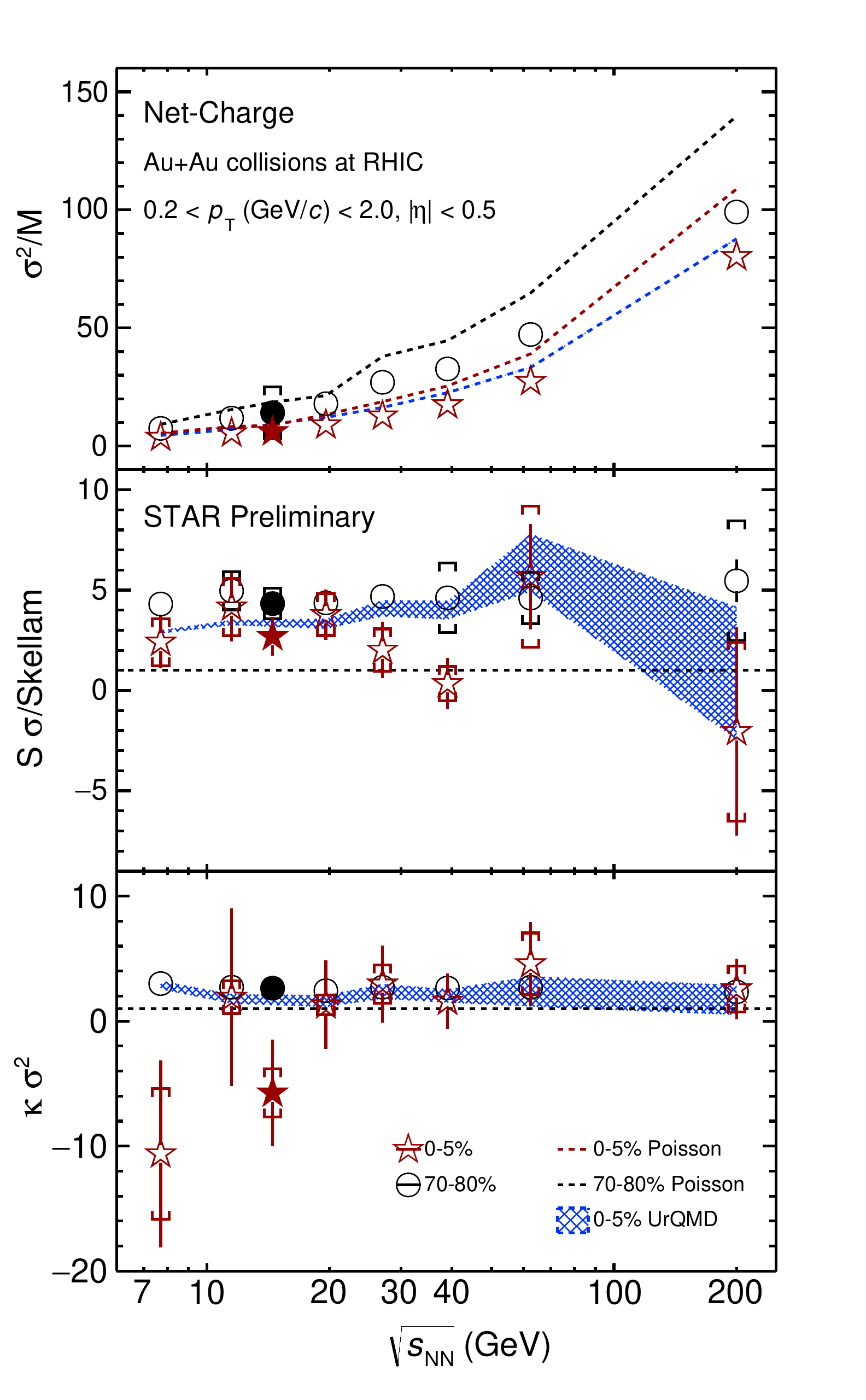}
\includegraphics[scale=0.30]{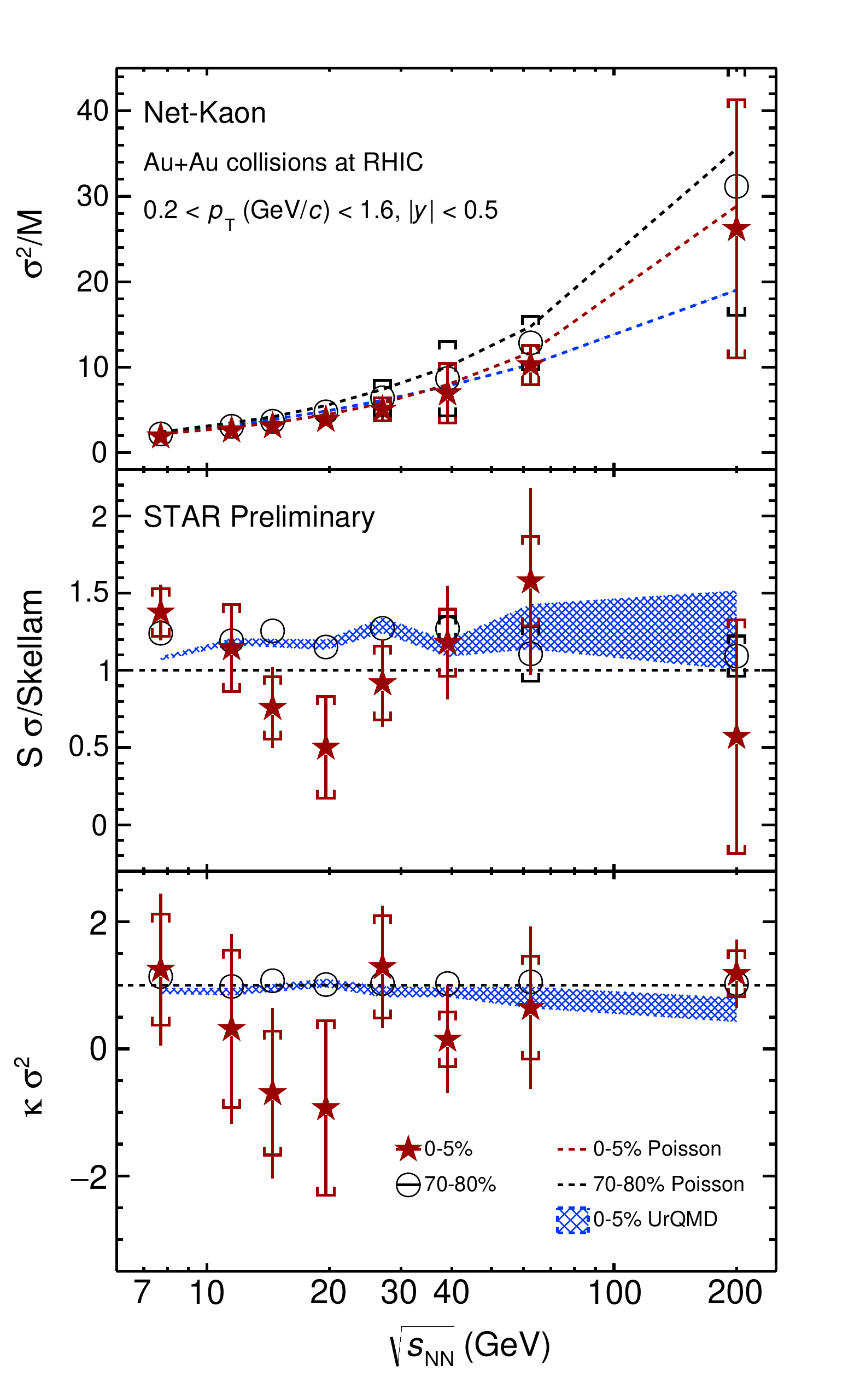}
\includegraphics[scale=0.30]{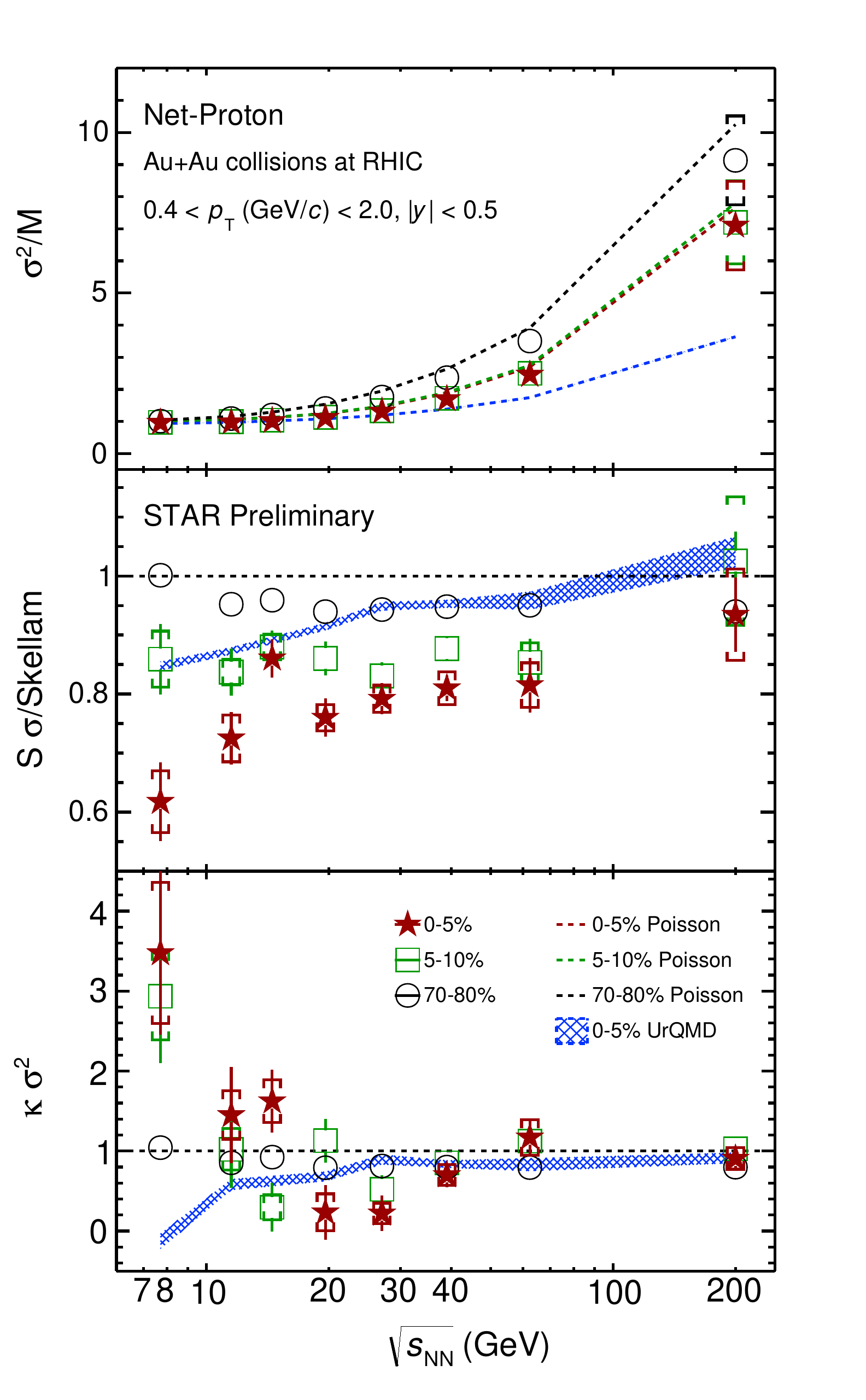}
\vspace*{-5mm}
\caption{Energy dependence of the volume independent cumulant ratios of the net-charge (left figure), net-kaon (middle figure), and net-proton (right figure) distributions. Showing \VM (top panels), \SDK (middle panels), and \KV (bottom panels) for top 0-5\% central (red stars), 5-10\% central (green squares), and 70-80\% peripheral (black circles) collisions. The poisson expectations are denoted as dotted lines and UrQMD calculations are shown as blue bands.}
\label{energyDependence}
\vspace*{-8mm}
\end{center}
\end{figure}

The collision energy dependence of the volume independent cumulant ratios \VM, \SDK, and \KV for top 0-5\% central, 5-10\% central, and 70-80\% peripheral collisions are presented in Fig.~\ref{energyDependence}, showing for the first time all fully corrected results from \netQ, \netK, and \netP distributions for all BES energies \snn = 7.7, 11.5, 14.5, 19.6, 27, 39, 62.4, and 200 \GeV. All results have been corrected for finite tracking and PID efficiencies, the statistical uncertainties have been calculated using the Delta Theorem, and the systematic uncertainties have been estimated by varying the track quality cuts and taking efficiency fluctuations into account \cite{Luo_error,PhysRevC.91.034907}. The full symbols on the left figure indicate the new results from \sqrtSnnE{14.5}, added to the already published net-charge results \cite{PhysRevLett.113.092301} (open symbols). No significant deviation from the Poisson expectation is observed within the uncertainties for \netQ and \netK cumulant ratios. Furthermore, the data are consistent within uncertainties with the UrQMD model calculations, which does not include a critical point. No energy dependence is observed for \SDK and \KV, as well as for the UrQMD calculations. 

However, a deviation from the Poisson expectation, as well as from the UrQMD calculations, is observed for \SDK and \KV of the net-proton distribution. A non-monotonic behavior of the net-proton \KV can be seen in top 0-5\% central collisions, with an increase at lower energies while peripheral collisions show much smaller deviations from Poisson statistics. UrQMD calculations show suppression at lower energies due to baryon number conservation. 

\begin{figure}[tbp]
\begin{center}
\vspace*{-3mm}
\includegraphics[scale=0.30]{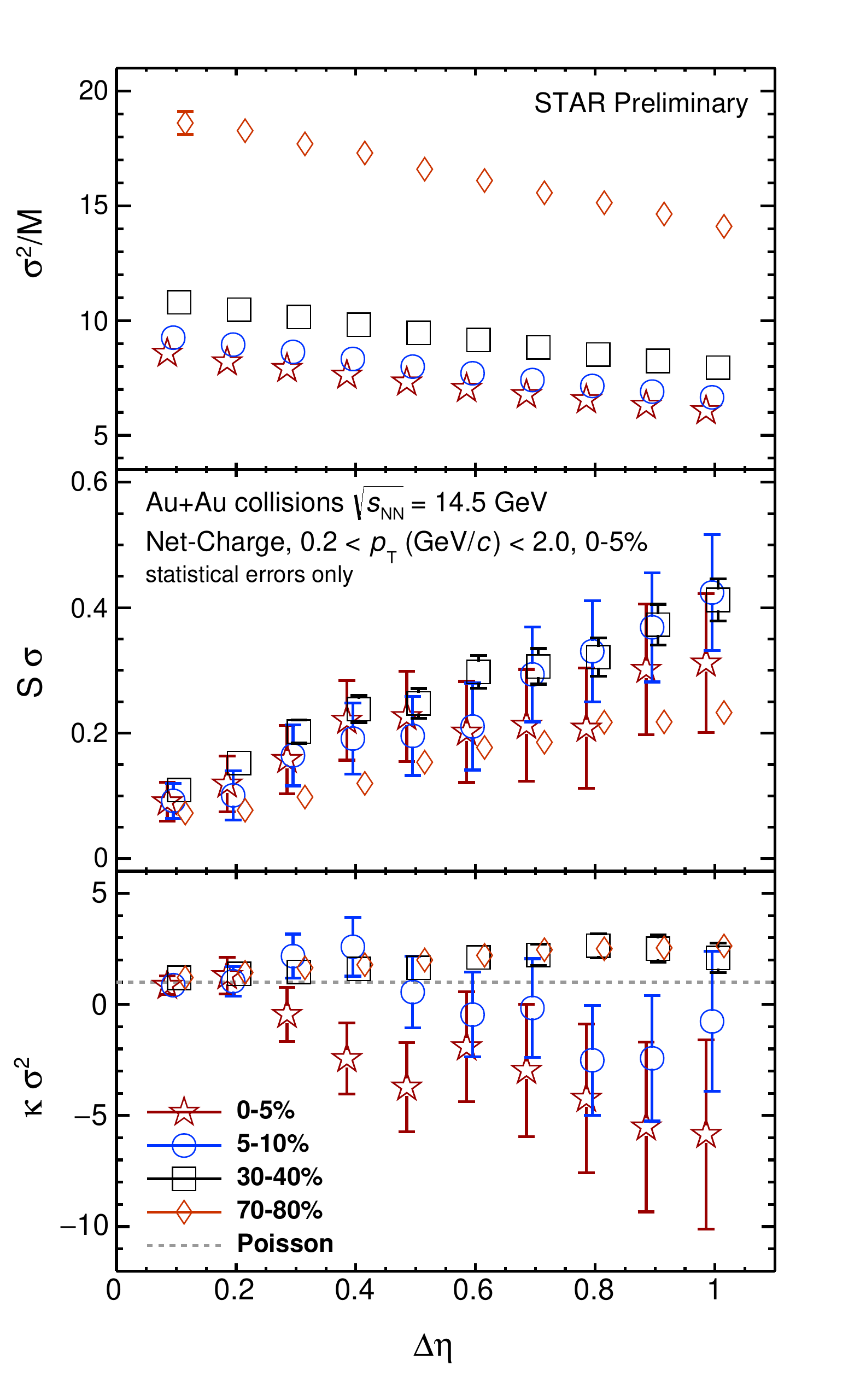}
\includegraphics[scale=0.30]{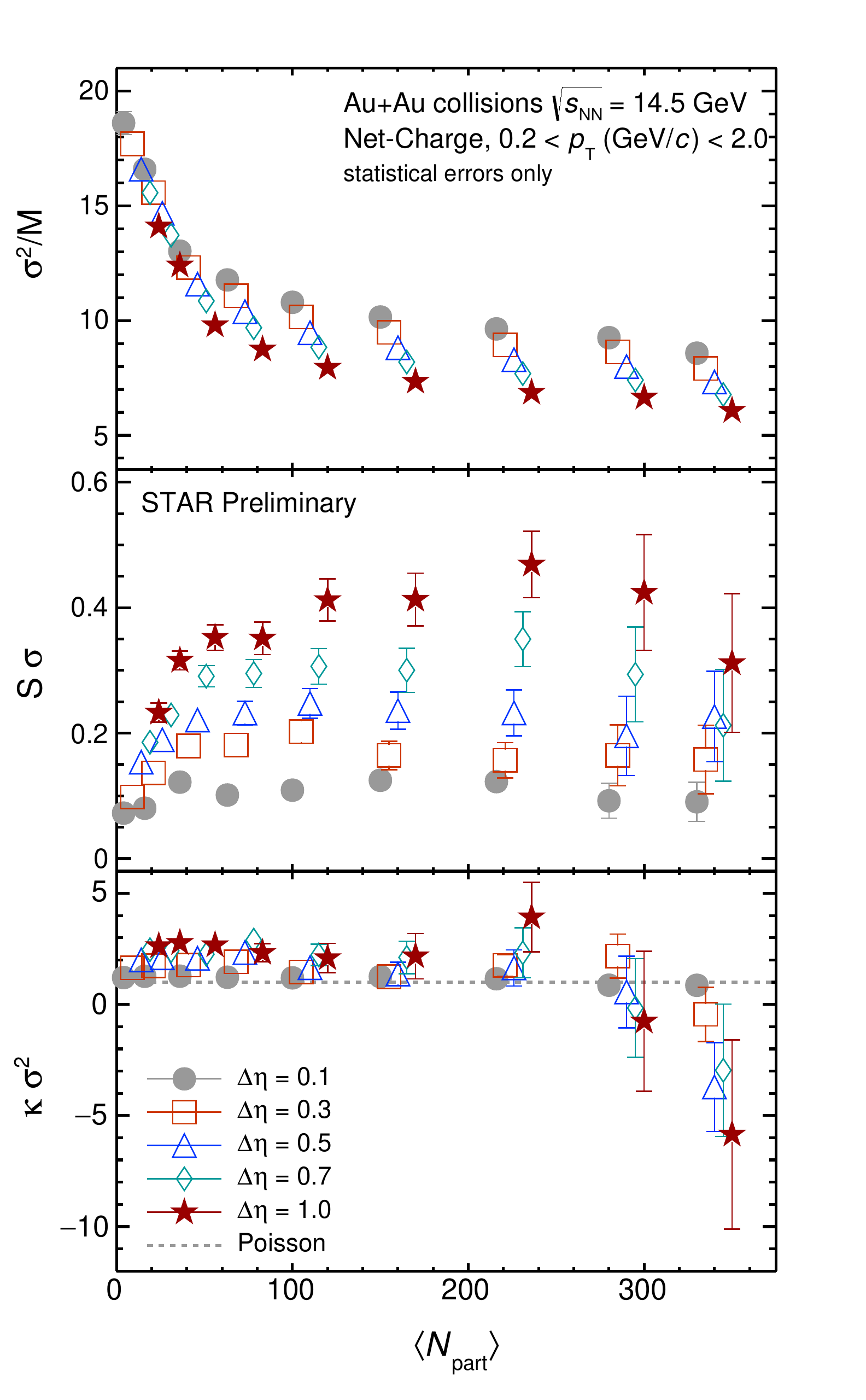}
\includegraphics[scale=0.26]{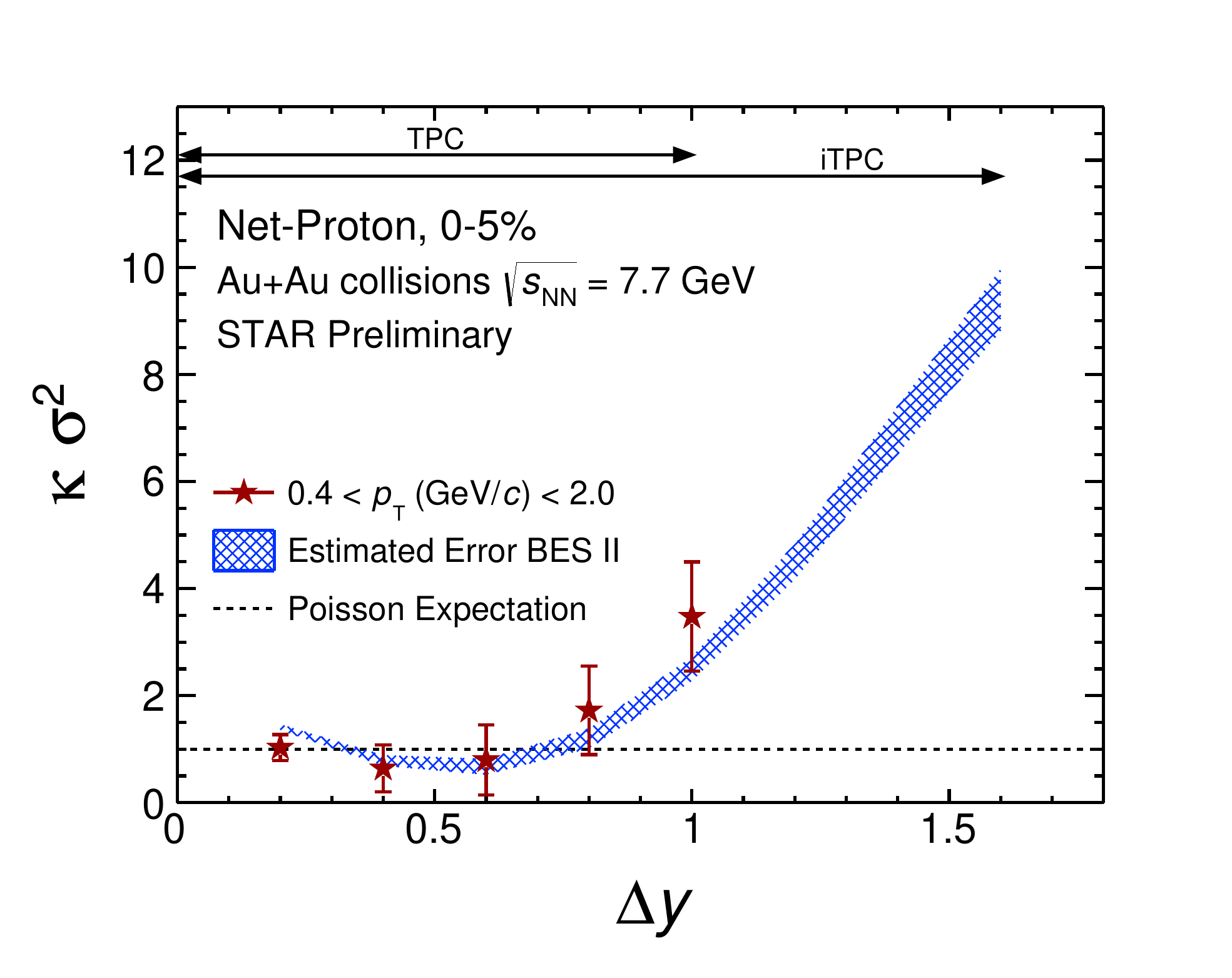}
\vspace*{-8mm}
\caption{Left two figures: pseudo-rapidity dependence of the volume independent cumulant ratios of the \netQ distributions at \sqrtSnnE{14.5}.
Right figure: error estimation (blue band) for net-proton \KV in Au+Au collisions at \sqrtSnnE{7.7} with the upgraded STAR detector and larger statistics in the upcoming RHIC Beam Energy Scan II for an increasing rapidity coverage.} 
\vspace*{-8mm}
\label{etaDependence}
\label{BESIIerror}
\end{center}
\end{figure}

The $\eta$ coverage (\deta) dependence of fluctuation observables allows the study of the history of the hot medium \cite{PhysRevC.90.064911,Kitazawa:2015aa}. Figure~\ref{etaDependence} shows the dependence of volume independent cumulant ratios of the \netQ distributions at \sqrtSnnE{14.5} on the pseudo-rapidity window width for different centralities (left figure), as well as, for different number of participants \meanNpart (middle figure). An ordering in \deta, as well as a smooth trend for \VM, \SD, and \KV with increasing \deta is observed. However, there seems to be a different tendency of \KV for central and peripheral collisions versus \deta. The smaller the \deta window, the closer to the Poisson expectation are the cumulant ratios \cite{Jeon:2003gk}, highlighting the need for a larger detector acceptance to make the most sensitive measurements.

The upcoming RHIC BES-II in 2019-2020 will include an upgraded STAR detector. An ``iTPC" upgrade will enlarge the phase-space up to $|\eta|<1.5$ and down to \pt = \unit[60]{\MeVc} and the Event-Plane Detector at forward rapidities will allow for a better centrality estimation, suppressing auto-correlations. Figure~\ref{BESIIerror} (right) shows the estimated statistical error for the net-proton \KV at \sqrtSnnE{7.7} with an expected increased signal, highlighting the importance of the enlarged rapidity coverage for the critical point search.

\section{Conclusions}
We present new results on cumulant ratios for \sqrtSnnE{14.5} Au+Au collisions for net-charge and net-proton distributions, completing the BES-I program for \sqrtSnnBES{7.7}{200}, as well as, for the first time, the fully corrected energy and centrality dependence of net-kaon cumulant ratios for all BES energies. Within uncertainties (statistical and systematic), net-charge and net-kaon results follow the Poisson expectation while a non-monotonic behavior is observed in the energy dependence of the net-proton \KV. Studying the pseudo-rapidity dependence of cumulant ratios shows that enlarging the detector acceptance is crucial for the critical point search with the STAR detector. The RHIC BES-II will bring a larger event sample and wider phase-space to be more sensitive for the search for critical behavior.





\bibliographystyle{elsarticle-num}
\bibliography{Moments_mod}







\end{document}